\documentclass[conference]{IEEEtran}

\usepackage[utf8]{inputenc}
\usepackage{graphicx}
\usepackage{amssymb}
\usepackage[cmex10]{amsmath}
\usepackage{color}
\usepackage{theorem}
\usepackage{pifont}
\usepackage{hyperref}
\usepackage{url}
\usepackage{breakurl}

\usepackage{bbm}

\usepackage[ruled,vlined,titlenumbered]{algorithm2e}
\newtheorem{Theorem}{Theorem}

\newtheorem{Lemma}{Lemma}
\newtheorem{Problem}{Problem}

\DeclareMathOperator{\rsgcd}{rsgcd}

\DeclareMathOperator{\rank}{rank}
\newcommand{\qed}{\hfill \mbox{\raggedright \rule{.07in}{.1in}}}
\IEEEoverridecommandlockouts

\begin{document}

\title{A Basis for all Solutions of the Key Equation for Gabidulin Codes}

\author{\IEEEauthorblockN{Antonia Wachter, Vladimir Sidorenko and Martin Bossert}
\IEEEauthorblockA{Institute of Telecommunications and Applied Information Theory\\
University of Ulm, Germany\\
\texttt{\{antonia.wachter | vladimir.sidorenko | martin.bossert\}@uni-ulm.de}}
\thanks{This work was supported by the German Research Council "Deutsche Forschungsgemeinschaft" (DFG) under Grant No. Bo867/21-1.}}
\maketitle
%\IEEEpeerreviewmaketitle

\begin{abstract}
We present and prove the correctness of an efficient algorithm that provides a basis for all solutions of a key equation in order to decode Gabidulin ($\mathcal{G}$-) codes up to a given radius $\tau$. This algorithm is based on a symbolic equivalent of the \textit{Euclidean Algorithm} (EA) and can be applied for decoding of $\mathcal{G}$-codes beyond half the minimum rank distance. %We give a criterion to check if the key equation has a unique solution. 
If the key equation has a unique solution, our algorithm reduces to Gabidulin's decoding algorithm up to half the minimum distance. If the solution is not unique, we provide a basis for all solutions of the key equation. Our algorithm has time complexity $\mathcal O(\tau^2)$ and is a generalization of the modified EA by Bossert and Bezzateev for Reed-Solomon codes.
\end{abstract}

% \begin{IEEEkeywords}
% 
% \end{IEEEkeywords}

\section{Introduction}
A special class of rank-metric codes was introduced by Delsarte \cite{Delsarte_1978}, Gabidulin \cite{Gabidulin_TheoryOfCodes_1985} and Roth \cite{Roth_RankCodes_1991}. These codes are also called Gabidulin ($\mathcal{G}$-) codes. K\"otter and Kschischang recently constructed network codes based on $\mathcal{G}$-codes \cite{koetter_kschischang}.

In \cite{Gabidulin_TheoryOfCodes_1985}, Gabidulin presented an algorithm for decoding $\mathcal{G}$-codes up to half the minimum (rank) distance with a symbolic equivalent of the \textit{Euclidean Algorithm} (EA). Paramonov and Tretjakov \cite{Paramonov_Tretjakov_BMA_1991} and independently Richter and Plass \cite{Richter_RankCodes_2004}, \cite{RichterPlass_DecodingRankCodes_2008} gave a generalization of the \textit{Berlekamp-Massey Algorithm} (BMA) for decoding of $\mathcal{G}$-codes up to half the minimum distance. This algorithm was proved and extended by Sidorenko \textit{et al.} in \cite{Sidorenko_ShiftRegister_2010}. This generalization of the BMA yields a basis for all solutions of the key equation for decoding of $\mathcal{G}$-codes up to a given radius, if there is no unique solution.

%Bossert and Bezzateev (BB) presented in \cite{IRS_Bossert} a method of decoding (interleaved) RS codes beyond half the minimum distance using the EA. 

In this paper, we present an algorithm that provides a basis for all solutions of the key equation using the symbolic equivalent of the EA. Our algorithm is a generalization of the \textit{Bossert-Bezzateev Algorithm} (BBA) from \cite{IRS_Bossert}. The BBA was applied for decoding of interleaved \textit{Reed-Solomon} (RS) codes beyond half the minimum distance using the EA. 

This paper is organized as follows: In Section~\ref{sec:defi}, we give the required definitions and state the problem. Section~\ref{sec:algo} provides the algorithm and in Section~\ref{sec:proofs}, we prove the correctness of the algorithm. The paper ends with a conclusion in Section~\ref{sec:concl}.

\section{Definitions and Problem Formulation}\label{sec:defi}
\subsection{Linearized Polynomials}
$\mathcal{G}$-codes are defined by means of \textit{linearized polynomials} (see e.g. \cite{Lidl-Niederreiter:FF1996}). %, \cite{Roth_RootsLinPoly}). 
Let $q$ be a power of a prime and let us denote the Frobenius $q$-power by:
\begin{equation}
x^{[i]}= x^{q^i}.
\end{equation}
A linearized polynomial over $\mathbb{F}_{q^m}$ is a polynomial of the form
\begin{equation}
L(x) = \sum \limits_{i=0}^{t} l_i x^{[i]},
\end{equation}
with $l_i \in \mathbb{F}_{q^m}$. %The \textit{q-degree} of $L(x)$ is defined by $\deg_q (L(x)) = t \Leftrightarrow \deg(L(x)) = q^t$.
If the coefficient $l_t\neq 0$, we define the \textit{q-degree} by $\deg_q L(x) = t$. 

An important property of linearized polynomials for all $a,b \in \mathbb{F}_{q^m}$ and all $\beta_1,\beta_2 \in \mathbb{F}_{q}$ is:
\begin{equation}\label{eq:linpolyroots}
%\begin{split}
L(\beta_1 a+\beta_2 b) = \beta_1 L(a)+\beta_2 L(b). %\qquad \forall ,\forall \alpha,\beta \in GF(q)\\
% L(c\cdot a) &= c \cdot L(a), \qquad \forall a \in GF(q^m), \ \forall c \in GF(q),
% \end{split}
\end{equation} 
Consequently, any linear combination of roots of a linearized polynomial $L(x)$ is also a root of $L(x)$.

Let $F(x)$ and $G(x)$ be linearized polynomials over $\mathbb{F}_{q^m}$. The \textit{symbolic product} of $F(x)$ and $G(x)$ is:
\begin{equation}
F(x) \otimes G(x) = F(G(x)).
\end{equation}
If $\deg_q F(x)=t_F$ and $\deg_q G(x)=t_G$, then $\deg_q(F(x) \otimes G(x))=t_F+t_G$.
The symbolic product satisfies associativity and distributivity, but in general it is non-commutative, i.e.: $F(x) \otimes G(x) \neq G(x) \otimes F(x)$. 

We call $G(x)$ a \textit{right symbolic divisor} of $A(x)$, if $A(x) = F(x) \otimes G(x)$ for some linearized polynomial $F(x)$.
% An important property concerning the right symbolic divisor is given by the following Lemma:
% \begin{Lemma}
% \cite{Roth_RootsLinPoly} A linearized polynomial $G(x)$ is a right symbolic divisor of a linearized polynomial $F(x)$ over $GF(q^m)$ if and only if $G(x)$ divides $F(x)$ in the ordinary sense.
% \end{Lemma}
These operations convert the set of all linearized polynomials into a non-commutative ring with the identity element $x^{[0]}=x$.

% In addition, we define a \textit{symbolic matrix multiplication} of two matrices:
% \begin{equation}
% \mathbf C = \mathbf A \otimes \mathbf B,
% \end{equation}
% which is calculated in the same way as the ordinary matrix product, except that the elements are multiplied symbolically, i.e. the element $c_{i,j}$ in row $i$ and column $j$ of $\mathbf C$ is calculated by:
% \begin{equation}
% c_{i,j} = \sum\limits_{m=1}^{n} a_{i,m} \otimes b_{m,j},
% \end{equation}
% where $n$ is the number of columns of $\mathbf A$ and the number of rows of $\mathbf B$. Clearly, neither the order of the matrices nor the order of the symbolically multiplied elements is commutative.

We define a \textit{symbolic equivalent} of the \textit{Extended Euclidean Algorithm} (SEEA). Let $R_{-1}(x) = B(x)$ and $R_{0}(x) = A(x)$ be two linearized polynomials with $\deg_q B(x) > \deg_q A(x)$. The \textit{right symbolic greatest common divisor} (rsgcd) is calculated by the following recursion:
\begin{equation}\label{eq:seea}
\begin{split}
R_{-1}(x) &= Q_1(x) \otimes R_0(x) + R_1(x)\\
R_{0}(x) &= Q_2(x) \otimes R_1(x) + R_2(x)\\
&\dots\\
R_{j-2}(x) &= Q_{j}(x) \otimes R_{j-1}(x) + R_j(x)\\
R_{j-1}(x) &= Q_{j+1}(x) \otimes R_{j}(x),
\end{split}
\end{equation}
where $\deg_q R_i(x) < \deg_q R_{i-1}(x)$. The last non-zero remainder $R_{j}(x)$ is the $\rsgcd(A(x),B(x))$.

Let $U_i(x)$ and $V_i(x)$ be polynomials, which can be calculated recursively:
\begin{equation}\label{eq:calcuv}
\begin{split}
U_i(x) &= - Q_i(x) \otimes U_{i-1}(x) + U_{i-2}(x)\\
V_i(x) &= - Q_i(x) \otimes V_{i-1}(x) + V_{i-2}(x),
\end{split}
\end{equation}
with $U_{-1}(x)=\!0, U_0(x) =\! x^{[0]}$ and $V_{-1}(x)=x^{[0]}, V_0(x) = 0$. By means of these polynomials, we can write each remainder as a combination of the input polynomials $A(x)$ and $B(x)$:
\begin{equation}\label{eq:remohnemod}
R_i(x) = U_i(x) \otimes A(x) + V_i(x) \otimes B(x).
\end{equation}
An important property of the polynomials from the SEEA is:
\begin{equation}\label{eq:propseea}
\deg_q U_i(x) + \deg_q R_{i-1}(x)=\deg_q B(x).
\end{equation}
The proof of \eqref{eq:propseea} is similar to the proof for the usual EA \cite{Sugiyama_AMethodOfSolving_1975}.

\subsection{Gabidulin Codes and Their Key Equation}
A $\mathcal{G}$-codeword is a vector $\mathbf c \in \mathbb{F}_{q^m}^n$, where $n$ is the codeword length:
\begin{equation}
\mathbf c = (c_0, c_1, \dots, c_{n-1}).
\end{equation}
%with $c_i \in GF(q^m)$. 
This vector can be mapped on an $m \times n$ matrix $\mathbf C$ with entries from $\mathbb{F}_{q}$. The \textit{rank norm} $\rank_q(\mathbf c)$ is the rank of $\mathbf C$ over $\mathbb{F}_{q}$. %The \textit{rank weight} of a codeword is then $\wt(\mathbf c) = \rank(\mathbf c)$ and the \textit{minimum rank distance} $d$ is defined by:

For $n \leq m$, a linear $(n,k)$ $\mathcal{G}$-code over $\mathbb{F}_{q^m}$ is defined by its $(n-k) \times n$ parity check matrix $\mathbf H$ (see \cite{Gabidulin_TheoryOfCodes_1985}):
%A $(n,k,d=n-k+1)$ $\mathcal{G}$-code with $n \leq m$ is defined by its $(d-1) \times n$ parity check matrix $\mathbf H$ (see \cite{Gabidulin_TheoryOfCodes_1985}):
\begin{small}
\begin{equation}
\mathbf H =
\left( \begin{array}{cccc}
h_1 & h_2 & \dots& h_n\\
h_1^{[1]} & h_2^{[1]} & \dots& h_n^{[1]}\\
\vdots&\vdots&\vdots&\vdots\\
h_1^{[n-k-1]} & h_2^{[n-k-1]} & \dots& h_n^{[n-k-1]}\\
\end{array}
\right),
\end{equation}
\end{small}
where $h_1, \dots, h_n \in \mathbb{F}_{q^m}$ are linearly independent over $\mathbb{F}_{q}$. The \textit{minimum rank distance} $d$ of the code $\mathcal G$ is defined by:
\begin{equation}
d = \min \lbrace \rank_q(\mathbf c) \; | \; \mathbf c \in \mathcal G, \mathbf c \neq \mathbf 0 \rbrace,
\end{equation}
and is $d=n-k+1$.

The transmitted codeword $\mathbf c$ is corrupted by an additive error $\mathbf e$ with $e_i \in \mathbb{F}_{q^m}$ of $\rank_q(\mathbf e)=t$:
\begin{equation}
\mathbf r = \mathbf c+ \mathbf e,
\end{equation}
where $\mathbf r$ is the received vector with $r_i \in \mathbb{F}_{q^m}$. We use a $t \times n$ matrix $\mathbf Y$ of rank $t$ with elements from $\mathbb{F}_{q}$ to write:
\begin{equation}\label{eq:decompe}
\mathbf e = \mathbf E \cdot\mathbf Y = (E_1,E_2,\dots,E_{t})\cdot\mathbf Y,
\end{equation}
with $E_1,E_2,\dots,E_{t}$ from $\mathbb{F}_{q^m}$ are linearly independent over $\mathbb{F}_{q}$. 
The syndrome $\mathbf s$ is calculated by:
\begin{equation}\label{eq:calcs}
\mathbf s = \mathbf r \cdot \mathbf H^T = \mathbf e \cdot \mathbf H^T= (S_0, S_1, \dots, S_{d-2}),
\end{equation}
and can also be represented as a linearized polynomial:
\begin{equation}
S(x) = \sum \limits_{i=0}^{d-2} S_i x^{[i]}.
\end{equation}
%As for RS codes, we define a \textit{key equation} for $\mathcal{G}$-codes:
The most important part of decoding $\mathcal{G}$-codes is solving the following key equation in order to find an \textit{error span polynomial} $\Lambda(x)$ and an auxiliary polynomial $\Omega(x)$. $\Lambda(x)$ contains all linear combinations of $E_1,E_2,\dots,E_{t}$ \eqref{eq:decompe} as roots.
\vspace{-1ex}
\begin{Theorem}\cite{Gabidulin_TheoryOfCodes_1985} Let the syndrome $S(x)$ and the minimum distance $d$ be known, then the key equation for $\mathcal{G}$-codes is:
\begin{equation}\label{eq:keyequation}
\Omega(x) = \Lambda(x) \otimes S(x) \mod x^{[d-1]},
\end{equation}
where $\Omega(x)$ and $\Lambda(x)$ are linearized polynomials with the degree constraints: 
\begin{equation}\label{eq:degconstrke}
%\deg_q \Lambda(x) = \tau, \ \deg_q \Omega(x) < \tau.
\deg_q \Omega(x) < \deg_q \Lambda(x).
\end{equation}
\end{Theorem}
% $\Lambda(x)$ is called \textit{error span polynomial}. A proper $\Lambda(x)$ has $q$-degree $\tau$ and contains all linear combinations of $\tau$ linearly independent elements $E^*_1,E^*_2,\dots,E^*_{\tau}$ as roots. The true $\Lambda(x)$ contains all linear combinations of $E_1,E_2,\dots,E_{t}$ as roots, where $t \leq \tau$.
%The most important part of decoding $\mathcal{G}$-codes is solving the key equation. 
In this paper, we restrict ourselves to solving the key equation. The decoding procedure afterwards can be done as explained e.g. in \cite{Gabidulin_TheoryOfCodes_1985}. In the following, let $\tau$ denote the decoding radius.

\subsection{Solving the Key Equation up to Half the Minimum Distance}
Gabidulin presented in \cite{Gabidulin_TheoryOfCodes_1985} a method to solve the key equation using the SEEA if $\deg_q \Lambda(x) \leq \tau=\left\lfloor \frac{d-1}{2}\right\rfloor$. The Gabidulin algorithm is similar to the algorithm by Sugiyama \textit{et al.} for RS codes \cite{Sugiyama_AMethodOfSolving_1975} and is shown in Algorithm~\ref{algo:UpToHalfD}. In this procedure, the SEEA runs with the input polynomials $R_{-1}(x)=x^{[d-1]}$ and $R_0(x)=S(x)$ (Lines~\ref{line:seeastart1}-\ref{line:seeaend1}) until the degree constraints of Line~\ref{line:degconstr} are fulfilled. Line~\ref{line:lambda} yields the solution to the key equation, where $a \in \mathbb{F}_{q^m}$ is an arbitrary constant factor. Often, $a$ is chosen such that $\Lambda(x)$ is monic. %We are only interested in the roots of $\Lambda(x)$ and hence, a constant factor does not matter, but often $c$ is chosen such that $\Lambda(x)$ is monic.
Similar as for RS codes, it can be shown that the polynomials $\Lambda(x)$ and $\Omega(x)$ are unique except for the constant factor $a$.%, if $\tau \leq\left\lfloor \frac{d-1}{2}\right\rfloor$.

\begin{algorithm}[htp]\vspace{1ex}
\caption{Solving the Key Equation if $\tau = \left\lfloor \frac{d-1}{2}\right\rfloor$ \cite{Gabidulin_TheoryOfCodes_1985}}
\label{algo:UpToHalfD}
\dontprintsemicolon
\SetVline
%\BlankLine
\linesnumbered
\SetKwInput{KwIn}{\underline{Input}}
\SetKwInput{KwOut}{\underline{Output}}
\SetKwInput{KwIni}{\underline{Initialize}}
\KwIn{Syndrome $S(x)$, $x^{[d-1]}$}
\KwIni{$i \leftarrow 0$, $R_{-1}(x) \leftarrow x^{[d-1]}$, $R_0(x) \leftarrow S(x)$, $U_{-1}(x) \leftarrow 0$, $U_0(x)   \leftarrow x^{[0]}$}
\While{\nllabel{line:seeastart1}$R_{i}(x) \neq 0$}
{$i \leftarrow i + 1$\;
% Calculate $Q_i(x), R_i(x), U_i(x)$ such that:\;
% $\qquad R_{i-2}(x)=Q_i(x)\otimes R_{i-1}(x) + R_i(x)$\;
% $\qquad U_i(x) = - Q_i(x) \otimes U_{i-1}(x) + U_{i-2}(x)$\nllabel{line:seeaend1}\;
Calculate $Q_i(x)$ and $R_i(x)$ such that:\;
$\qquad R_{i-2}(x)=Q_i(x)\otimes R_{i-1}(x) + R_i(x)$\;
\BlankLine
$U_i(x) \leftarrow - Q_i(x) \otimes U_{i-1}(x) + U_{i-2}(x)$\nllabel{line:seeaend1}\;
\BlankLine
\If{\nllabel{line:degconstr}$\deg_q \!R_{i-1}(x) \! \geq \! \left\lfloor \frac{d-1}{2}\right\rfloor\! $ \textbf{and} $\!\deg_q\! R_{i}(x)  \!<  \!\left\lfloor \frac{d-1}{2}\right\rfloor\!$}
{\textbf{break}}
}
$\Lambda(x) \leftarrow a \cdot U_i(x)$ and $\Omega(x) \leftarrow a \cdot R_i(x)$\nllabel{line:lambda}\;
\KwOut{$\Lambda(x)$, $\Omega(x)$}\;
\vspace{1ex}
\end{algorithm}

Note that a proper $\Lambda(x)$ has $q$-degree $t_{\lambda}=\deg_q \Lambda(x) \leq \tau$ and contains all linear combinations of $t_{\lambda}$ linearly independent elements from $\mathbb{F}_{q^m}$ as roots. In practice, if Algorithm~\ref{algo:UpToHalfD} fails, decoding can be done with our algorithm that we provide in Section~\ref{sec:algo}. %Algorithm~\ref{algo:BeyondHalfD}. %with $E_1,E_2,\dots,E_{t}$ from $\mathbb{F}_{q^m}$ are linearly independent over $\mathbb{F}_{q}$

% We can recognize during Algorithm~\ref{algo:UpToHalfD} if $\rank_q(\mathbf e) <  \left\lfloor \frac{d-1}{2}\right\rfloor$ with the following lemma. The proof is skipped due to place restrictions.
% \vspace{-1ex}
% \begin{Lemma}\label{theo:ranksmall}
% $\rank_q(\mathbf e) < \left\lfloor \frac{d-1}{2}\right\rfloor$, if and only if there exists a pair $\lbrace R_i(x),U_i(x) \rbrace$ in the SEEA with $\deg_q R_i(x) < \deg_q U_i(x) < \left\lfloor \frac{d-1}{2}\right\rfloor$.
% \end{Lemma}

%In this paper, we will concentrate only on solving the Key Equation as this is the hard part of decoding and hence, we do not go into detail about the remaining decoding procedure.

\subsection{Problem Formulation}
The problem of finding all solutions of the key equation if $\tau >\left\lfloor \frac{d-1}{2}\right\rfloor$ can be formulated as follows.\vspace{-1ex}
\begin{Problem}\label{prob:dec}
Let an integer $\tau$, with $\left\lfloor \frac{d-1}{2}\right\rfloor<\tau<d-1$, the syndrome $S(x)$ and $x^{[d-1]}$ be known. Assume, $S(x)$ results from an error vector $\mathbf e$ with $\rank_q(\mathbf e) \leq \tau$ \eqref{eq:calcs}. Find all pairs of polynomials $\lbrace\Lambda(x),\Omega(x)\rbrace$ with 
\begin{equation}\label{eq:degprob}
\deg_q \Omega(x) < \deg_q \Lambda(x)=\tau,
\end{equation}
such that the key equation \eqref{eq:keyequation} 
% \begin{equation}
% \Omega(x) = \Lambda(x) \otimes S(x) \mod x^{[d-1]}
% \end{equation}
is fulfilled. %with 
% \begin{equation}\label{eq:degprob}
% \deg_q \Omega(x) < \deg_q \Lambda(x)=\tau.
% \end{equation}
\end{Problem}

We want to solve this problem using the SEEA in an efficient way. The restriction $\rank_q(\mathbf e) \leq \tau$ means that we limit the decoding radius to $\tau$. We introduce $\tau_0$:
\begin{equation}\label{eq:tauzero}
\tau=\Big\lfloor \frac{d-1}{2}\Big\rfloor+\tau_0.
\end{equation}

Of course, the key equation \eqref{eq:keyequation} can be solved with standard methods (e.g. Gaussian elimination) with complexity $\mathcal O(\tau^4)$ operations in $\mathbb{F}_{q^m}$. However, we want an \textit{efficient} solution. 

% If $\tau \leq \left\lfloor \frac{d-1}{2}\right\rfloor$, an efficient solution with complexity $\mathcal O(n^2)$ is given by Algorithm~\ref{algo:UpToHalfD}. In this case, the solution is unique. We now consider $\tau >\left\lfloor \frac{d-1}{2}\right\rfloor$, where the solution of Problem~\ref{prob:dec} is not unique and we want to obtain all solutions for Problem~\ref{prob:dec}. 
In the following, we give an efficient algorithm to solve Problem~\ref{prob:dec} that has complexity $\mathcal O(\tau^2)$. The main results of our paper are Algorithm~\ref{algo:BeyondHalfD} and Theorem~\ref{theo:linkombi}. %that provides the most important proof of the algorithm.

\section{The Algorithm for Solving the Key Equation}\label{sec:algo}
In this section, we give an efficient algorithm (Algorithm~\ref{algo:BeyondHalfD}) based on the SEEA which solves Problem~\ref{prob:dec} with complexity $\mathcal O(\tau^2)$. %if $\tau >\left\lfloor \frac{d-1}{2}\right\rfloor$. In this case, the solution is not unique and Algorithm~\ref{algo:BeyondHalfD} provides all solutions of Problem~\ref{prob:dec}. 
We explain the different steps of Algorithm~\ref{algo:BeyondHalfD} in this section and give the proofs in Section~\ref{sec:proofs}.

\begin{algorithm}[htp]%\vspace{1ex}
\label{algo:BeyondHalfD}
\caption{Basis for all Solutions if $\tau >\left\lfloor \frac{d-1}{2}\right\rfloor$}
\dontprintsemicolon
\SetVline
%\BlankLine
\linesnumbered
\SetKwInput{KwIn}{\underline{Input}}
\SetKwInput{KwOut}{\underline{Output}}
\SetKwInput{KwIni}{\underline{Initialize}}
\KwIn{Syndrome $S(x)$, $d$, $\tau$}
\KwIni{$i \!\leftarrow\! 0$, $j\! \leftarrow \!0$, $R_{-1}(x)\! \leftarrow\! x^{[d-1]}$, $R_0(x\!) \leftarrow\! S(x)$, $U_{-1}(x) \leftarrow 0$, $U_0(x) \leftarrow x^{[0]}$,\hspace{10ex} $\Delta_0(x) \leftarrow x^{[0]}$, $P_0(x) \leftarrow S(x)$}
\BlankLine
\While{\nllabel{line:seeastart}$R_{i}(x) \neq 0$}
{$i \leftarrow i + 1$, $j \leftarrow j + 1$\;
Calculate $Q_i(x)$ and $R_i(x)$ such that:\;
$\qquad R_{i-2}(x)=Q_i(x)\otimes R_{i-1}(x) + R_i(x)$\nllabel{line:calcremainder}\;
\BlankLine
$U_i(x) \leftarrow - Q_i(x) \otimes U_{i-1}(x) + U_{i-2}(x)$\nllabel{line:seeaend}\;
%Run SEEA until the end $\Rightarrow \lbrace R_i(x), U_i(x)\rbrace$\;
\While{\nllabel{line:deltasstart}$\deg_q U_i(x)-\deg_q \Delta_{j-1}(x)>1$}
{$\Delta_j(x) \leftarrow x^{[1]} \otimes \Delta_{j-1}(x)$\;
$P_j(x) \leftarrow x^{[1]} \otimes P_{j-1}(x)$\;
$j \leftarrow j + 1$\nllabel{line:deltasend}\;
}
$\Delta_j(x) \leftarrow U_i(x)\ $ and $\ P_j(x) \leftarrow R_i(x)$\;}
%Fill up missing degrees as in Def.~\ref{defi:sets} $\rightarrow \ \lbrace P_i(x), \Delta_i(x)\rbrace$\nllabel{line:deltas}\;
%Calculate set $\mathcal I$ \eqref{eq:setI} $\rightarrow \ \Delta^{\mathcal I}, P^{\mathcal I}$ \eqref{eq:deltaI} \nllabel{line:setI}\;
Calculate $\Delta^{\mathcal I}, P^{\mathcal I}$ using \eqref{eq:setI}, \eqref{eq:deltaI} \nllabel{line:setI}\;
% All solutions of Problem~\ref{prob:dec} are given by: $\Lambda(x)  = \sum \limits_{i=1}^{2\tau_0+1} \beta_i \Delta_i(x)$ and $\Omega(x)  = \sum \limits_{i=1}^{2\tau_0+1} \beta_i P_i(x)$\;\vspace{1ex}
% \KwOut{$\lbrace \Lambda(x), \ \Omega(x)\rbrace$}\;
\BlankLine
\KwOut{$\Delta^{\mathcal I}, P^{\mathcal I}$}
\vspace{1ex}
\end{algorithm}

Algorithm~\ref{algo:BeyondHalfD} executes the SEEA \eqref{eq:seea} with the input polynomials $R_{-1}(x) = x^{[d-1]}$ and $R_0(x)=S(x)$. In each step $i$, the SEEA returns the remainder $R_i(x)$ and the polynomial $U_i(x)$ \eqref{eq:calcuv}. This corresponds to Lines~\ref{line:seeastart} until \ref{line:seeaend} of Algorithm~\ref{algo:BeyondHalfD}. 

The $q$-degree of the remainder $R_i(x)$ is decreasing in every step, but not necessarily by one. Equivalently, the $q$-degree of $U_i(x)$ is increasing in every step of the SEEA, but also not necessarily by one. Hence,
\begin{equation}
\begin{split}
\deg_q R_{i+1}(x) &\leq \deg_q R_i(x)-1,\\
\deg_q U_{i+1}(x) &\geq \deg_q U_i(x)+1.
\end{split}
\end{equation}
% In order to obtain polynomials with all $q$-degrees from $0$ to $d-2$, we fill up the missing degrees in the pairs of polynomials $\lbrace U_i(x), R_i(x)\rbrace$ (Lines~\ref{line:deltasstart}-\ref{line:deltasend}). Definition~\ref{defi:sets} provides a rule for filling up these gaps in the $q$-degrees.
% \vspace{-1ex}
% \begin{Definition}\label{defi:sets}
% Let us define new pairs of polynomials $\lbrace\Delta_i(x),P_i(x)\rbrace$ based on $\lbrace U_i(x),R_i(x)\rbrace$. If a certain $q$-degree of $U_i(x)$ is missing, we fill it up by symbolically multiplying the previous polynomials by $x^{[1]}$ from the left for all $j=0,\dots,d-2$:
% \vspace{-1ex}\\
% \begin{itemize}
% \item $\Delta_0(x) = U_0(x)=x^{[0]} \ $ and $\ P_0(x) = A(x)=S(x)$,\\[-1.5ex]
% \item if $\ \exists \ U_i(x): \deg_q U_i(x) = j$, then: \\  \mbox{$\Delta_j(x) = U_i(x)\ $ and $\ P_j(x) = R_i(x)$},\\[-1.5ex]
% \item if $\ \nexists \ U_i(x): \deg_q U_i(x) = j$, then: \\  \mbox{$\Delta_j(x) = x^{[1]} \otimes \Delta_{j-1}(x)\ $ and $\ P_j(x) = x^{[1]} \otimes P_{j-1}(x)$}.
% \end{itemize}
% \end{Definition}
%A gap in the $q$-degrees of $U_i(x)$ occurs in step $i$ if the degree of the remainders decreases by $\delta > 1$ in the previous step $i-1$. Therefore, in both sets of $U_i(x)$ and $R_i(x)$, there are $\delta-1$ $q$-degrees missing. 
The missing degrees are then filled up as given in Lines~\ref{line:deltasstart} until \ref{line:deltasend} of Algorithm~\ref{algo:BeyondHalfD}. This definition assures that there exist polynomials $\Delta_i(x)$, $P_i(x)$ of each $q$-degree from $0$ to $d-2$. This is shown in Lemma~\ref{theo:linind}.

% \begin{Lemma}\label{theo:linind}
% All polynomials $\lbrace\Delta_i(x)\rbrace$ are linearly independent and all polynomials $\lbrace P_i(x) \rbrace$ are linearly independent.
% \end{Lemma}
\begin{Lemma}\label{theo:linind}
There exist polynomials of each degree from $0$ to $d-2$ in each set $\lbrace\Delta_i(x)\rbrace$ and $\lbrace P_i(x) \rbrace$.
\end{Lemma}
\begin{IEEEproof}
%If polynomials have different $q$-degrees, they are linearly independent. %This can be seen, when writing the coefficients of the polynomials as vectors. 
%Hence, it is sufficient to show that all $\Delta_i(x)$ and all $P_i(x)$ have different $q$-degrees. It does not matter if any $\Delta_i(x)$ is linearly dependent of any $P_i(x)$.
At first assume that $\deg_q U_{i+1}(x)=\deg_q U_{i}(x)+1$ and $\deg_q R_{i+1}(x)=\deg_q R_i(x) -1$ for all $i$. Hence, all $U_i(x)=\Delta_i(x)$ and all $R_i(x)=P_i(x)$ have different degrees.

Now, consider the case when some degrees have to be filled up in Lines~\ref{line:deltasstart} until \ref{line:deltasend}. Assume, that the $q$-degree of the $(i-1)$th remainder decreases by more than one, i.e.:
\begin{equation}
\deg_q R_{i-1}(x)=\deg_q R_{i-2}(x)-\delta \quad \text{with} \quad \delta>1.
\end{equation}
%, we obtain in step $i$:
% \begin{equation}
% \begin{split}
% R_{i-1}(x)&=Q_{i+1}(x) \otimes R_i(x) + R_{i+1}(x),\\
% U_{i+1}(x)&=-Q_{i+1}(x)\otimes U_i(x) + U_{i-1}(x).
% \end{split}
% \end{equation}
%With the calculation of $U_i(x)$ (Line~\ref{line:seeaend}) and 
Since $\deg_q R_{i}(x) < \deg_q R_{i-1}(x)$, we know using Line~\ref{line:calcremainder} from Algorithm~\ref{algo:BeyondHalfD}:
\begin{equation}
\deg_q Q_{i}(x)=\deg_q R_{i-2}(x)-\deg_q R_{i-1}(x)=\delta.
\end{equation}
With the calculation of $U_i(x)$ (Line~\ref{line:seeaend}), we obtain:
\begin{equation}
\deg_q U_{i}(x)=\deg_q U_{i-1}(x)+\deg_q Q_{i}(x)=\deg_q U_{i-1}(x)+\delta.
\end{equation}
Hence, if the degree of the remainders decreases by $\delta$ in step $i$, then the degree of $U_{i+1}(x)$ increases by $\delta$ in step $i+1$. If we  multiply $R_i(x)$ symbolically from the left with $x^{[1]}$ and do this $(\delta-1)$ times, we fill up the $\delta-1$ missing degrees of the remainders. The same holds for the $U_i(x)$. % and we see that the degree of the $\Delta_i(x)$ is increasing by one in each step. 

Due to \eqref{eq:propseea}, the $q$-degree of the last calculated $U_i(x)$ is $d-2$ and the $q$-degree of the last remainder is $0$. Thus, all degrees from $0$ to $d-2$ exist in the sets $\lbrace\Delta_i(x)\rbrace$ and $\lbrace P_i(x) \rbrace$. % have different degrees. %and hence are sets with linearly independent polynomials. %and hence $\Delta^{\mathcal I}$ and $P^{\mathcal I}$ are two sets of linearly independent polynomials (within these sets).
\end{IEEEproof}
\vspace{1.5ex}

If polynomials have different $q$-degrees, they are linearly independent. This becomes clear, if we write the coefficients of the polynomials as vectors. Consequently, with Lemma~\ref{theo:linind}, the polynomials $\lbrace \Delta_i(x) \rbrace$ and $\lbrace P_i(x) \rbrace$ are linearly independent. %The polynomials $\lbrace P_i(x) \rbrace$ are also linearly independent. %We prove this in Theorem~\ref{theo:linind}. 

\begin{figure}[htp]
\centering
\includegraphics[width=0.44\textwidth]{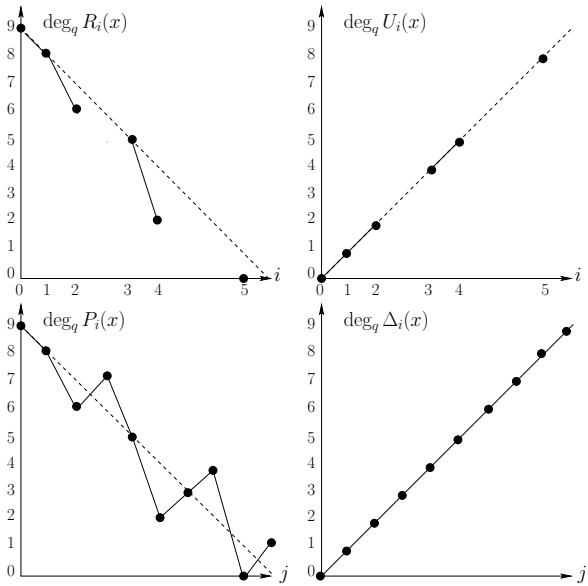}
\caption{Filling up the gaps in the degrees ($d=11$)}
\vspace{-2ex}
\label{fig:degrees}
\end{figure}

Figure~\ref{fig:degrees} shows an example how the gaps are filled up. The upper half of this figure shows the $q$-degrees of $R_i(x)$, $U_i(x)$ and the lower half the $q$-degrees of $P_i(x)$, $\Delta_i(x)$. For example, there is a $q$-degree difference of 2 between $R_1(x)$ and $R_2(x)$. Consequently, in the third step, $\deg_q U_3(x)=\deg_q U_2(x)+2$. The lower half of the picture shows that after filling up these $q$-degrees, all $q$-degrees from $0$ to $d-2$ exist.

Subsequently, we define a subset of these polynomials (Line~\ref{line:setI} of Algorithm~\ref{algo:BeyondHalfD}):
\begin{equation}\label{eq:setI}
\mathcal I = \lbrace i \;|\;\deg_q \Delta_i(x) \leq \tau \wedge \deg_q P_i(x) < \tau\rbrace.
\end{equation}
In Lemma~\ref{theo:numbPoly}, we show that the set $\mathcal I$ determines $2\tau_0+1$ pairs of polynomials $\lbrace\Delta_i(x),P_i(x)\rbrace$, which we denote by:
\begin{equation}\label{eq:deltaI}
\Delta^{\mathcal I}=\lbrace \Delta_i(x)\,|\,i \in \mathcal I\rbrace \quad \text{and} \quad P^{\mathcal I}=\lbrace P_i(x)\,|\,i \in \mathcal I\rbrace.
\end{equation}
%The solution of Problem~\ref{prob:dec} is not unique if  $\tau >\left\lfloor \frac{d-1}{2}\right\rfloor$. 
The following linear combinations of polynomials provide all solutions of Problem~\ref{prob:dec} (see Theorem~\ref{theo:linkombi}):
\begin{equation}\label{eq:linkombi}
\Lambda(x)  = \sum \limits_{i} \beta_i \Delta^{\mathcal I}_i(x) \quad \text{and} \quad
\Omega(x)  = \sum \limits_{i} \beta_i P^{\mathcal I}_i(x),
\end{equation}
% \begin{gather}\label{eq:linkombi}
% \Lambda(x)  = \sum \limits_{i} \beta_i \Delta^{\mathcal I}_i(x),\\
% \Omega(x)  = \sum \limits_{i} \beta_i P^{\mathcal I}_i(x),\label{eq:linkombi2}
% \end{gather}
with $\beta_i \in \mathbb{F}_{q^m}$. The definition of $\mathcal I$ assures that $\deg_q \Lambda(x) = \tau$ and $\deg_q \Omega(x) < \tau$. Lemma~\ref{theo:linkombKE} gives the proof that these linear combinations satisfy the key equation \eqref{eq:keyequation}. In Theorem~\ref{theo:linkombi} we prove that there are only $(q^m)^{2\tau_0+1}$ solutions of Problem~\ref{prob:dec} and that these linear combinations provide all solutions. 

Thus, Algorithm~\ref{algo:BeyondHalfD} provides a basis for all solutions of Problem~\ref{prob:dec}. The algorithm only requires a complexity of $\mathcal O(\tau^2)$ operations in $\mathbb{F}_{q^m}$ and is a generalization of the BBA \cite{IRS_Bossert}. The two algorithms are equivalent if $m=1$, i.e. if $\mathbb{F}_{q^m}=\mathbb{F}_{q}$, since then $\alpha ^{[i]}=\alpha$ for all elements from $\mathbb{F}_{q}$. 

% Note, that a proper $\Lambda(x)$ for $\mathcal{G}$-codes is defined as a linearized polynomial of $q$-degree $\tau$ with $\tau$ linearly independent roots over $\mathbb{F}_{q}$. If we use Algorithm~\ref{algo:BeyondHalfD} for decoding of $\mathcal{G}$-codes beyond half the minimum distance, we are only interested in the \textit{roots} of $\Lambda(x)$. One $\beta_i$ can be chosen freely as there are only $(q^m)^{2\tau_0}$ solutions with different roots.  %Of course, all linear combinations of these $\tau$ roots are also roots (see \eqref{eq:linpolyroots}). 
% Only a small subset of the $(q^m)^{2\tau_0}$ solutions provides proper $\Lambda(x)$. 

\section{Proof of the correctness of the Algorithm}\label{sec:proofs}
% This section provides the proof of Algorithm~\ref{algo:BeyondHalfD}. In Section~\ref{sec:algo} we already mentioned these proofs at the corresponding steps of the algorithm.

In order to prove that Algorithm~\ref{algo:BeyondHalfD} solves Problem~\ref{prob:dec}, we show with Lemmas~\ref{theo:linind} and \ref{theo:numbPoly} that each of the sets $\Delta^{\mathcal I}$ and $P^{\mathcal I}$ consists of $2\tau_0+1$ linearly independent polynomials. Afterwards, we prove that the linear combinations from \eqref{eq:linkombi} fulfill the key equation (Lemma~\ref{theo:linkombKE}) and that they give \textit{all} solutions of Problem~\ref{prob:dec} (Theorem~\ref{theo:linkombi}). Lemma~\ref{theo:rankS} gives some properties that we need for the proof of Theorem~\ref{theo:linkombi}.

\vspace{-1ex}
\begin{Lemma}\label{theo:numbPoly}
The set $\mathcal I=\lbrace i \;|\;\deg_q \Delta_i(x) \leq \tau \wedge \deg_q P_i(x) < \tau\rbrace$ \eqref{eq:setI} has cardinality $2\tau_0+1$. 
\end{Lemma}
\begin{IEEEproof}
At first assume, that $\deg_q U_{i+1}(x)=\deg_q U_{i}(x)+1$ and $\deg_q R_{i+1}(x)=\deg_q R_i(x) -1$ for all $i$. With \eqref{eq:propseea}:
\begin{equation}\label{eq:deguiri}
\deg_q U_i(x) + \deg_q R_i(x) = \deg_q B(x)-1=d-2.
\end{equation}
%$|\mathcal I|$ is determined by restricting the $q$-degrees of the pairs of polynomials. 
The remainder $R_m(x)$ with $\deg_q R_m(x)=\tau-1$ determines the smallest $i$ in the set $\mathcal I$ as $\deg_q R_i(x)$ is decreasing with increasing $i$. The polynomial $U_n(x)$ with $\deg_q U_n(x)=\tau$ determines the largest $i$ as $\deg_q U_i(x)$ is increasing with $i$. If $\deg_q U_n(x) = \tau$, we know with \eqref{eq:deguiri} that $\deg_q R_n(x) = d-2-\tau$. Hence, with \eqref{eq:tauzero}:
\begin{equation}
%\begin{split}
|\mathcal I|=\deg_q R_m(x)-\deg_q R_n(x)+1=2\tau_0+1.
%&=(\tau-1)-(d-\tau-2)+1=2\tau_0+1.
%\end{split}
\end{equation}

If some degrees have to be filled up, we know from Lemma~\ref{theo:linind} that the sets contain all degrees. Therefore, the number of polynomials in $\mathcal I$ does not change, only the elements change and the cardinality is also $2\tau_0+1$.
\end{IEEEproof}

Thus, with Lemma~\ref{theo:linind} and \ref{theo:numbPoly}, each set $\Delta^{\mathcal I}$ and $P^{\mathcal I}$ consists of $2\tau_0+1$ linearly independent polynomials.
%\vspace{2ex}
%Lemma~\ref{theo:rankS} gives some properties that we need for the proof of Theorem~\ref{theo:linkombi}.

\begin{Lemma}\label{theo:linkombKE}
Let $\Delta^{\mathcal I}$ and $P^{\mathcal I}$ be the sets of polynomials calculated by Algorithm~\ref{algo:BeyondHalfD}. Any pair of polynomials $\lbrace\Lambda(x),\Omega(x)\rbrace$ calculated by the linear combinations from \eqref{eq:linkombi} satisfies the key equation \eqref{eq:keyequation} with the degree constraints from \eqref{eq:degprob}.
% \begin{equation}\label{eq:keyeqtheo}
% \Omega(x) = \Lambda(x) \otimes S(x) \mod x^{[d-1]},
% \end{equation}
% with
% \begin{equation}\label{eq:keyeqdeg}
% \deg_q \Omega(x) < \tau, \ \deg_q \Lambda(x) \leq \tau.
% \end{equation}
\end{Lemma}
\begin{IEEEproof}
For each $R_i(x)$ and $U_i(x)$, the following holds:% (compare \eqref{eq:remohnemod}):
\begin{equation}
R_i(x) = U_i(x) \otimes S(x) \mod x^{[d-1]},
\end{equation}
hence every polynomial that is a direct output of the SEEA satisfies the key equation. For the polynomials that are filled up as in Algorithm~\ref{algo:BeyondHalfD}, we obtain:
\begin{equation}
x^{[k]} \otimes R_i(x) = x^{[k]} \otimes U_i(x) \otimes S(x) \mod x^{[d-1]},
\end{equation}
where $k$ is the number of missing degrees between $i$ and the current $q$-degree $j=i+k$. This is equivalent to:
\begin{equation}\label{eq:deltaPKE}
\Delta_{j}(x) = P_{j}(x) \otimes S(x) \mod x^{[d-1]}.
\end{equation}
%If we multiply both sides of \eqref{eq:deltaPKE} with $\beta_i \in \mathbb{F}_{q^m}$ and sum up over some $i$, we obtain due to the distributivity of the symbolic product:
Calculating the linear combinations from \eqref{eq:linkombi}, we obtain due to the distributivity of the symbolic product:
% \begin{equation}
% \sum \limits_{i} \beta_i\Delta^{\mathcal I}_{i}(x) = \sum \limits_{i} \beta_i P^{\mathcal I}_{i}(x) \otimes S(x) \mod x^{[d-1]}.
% \end{equation}
%Due to the distributivity of the symbolic product, we can factor out $\otimes S(x)$ on the right hand side and obtain:
\begin{equation}
\sum \limits_{i} \beta_i\Delta^{\mathcal I}_{i}(x) = \Big[\sum \limits_{i} \beta_i P^{\mathcal I}_{i}(x)\Big] \otimes S(x) \mod x^{[d-1]},
\end{equation}
which satisfies the key equation \eqref{eq:keyequation}. Due to the definition of $\mathcal I$ from \eqref{eq:setI}, also the degree constraints from \eqref{eq:degprob} are fulfilled.
\end{IEEEproof}
%The proof of Lemma~\ref{theo:linkombKE} has to be skipped due to place restrictions.
% We know from Theorem~\ref{theo:linind} and Lemma~\ref{theo:numbPoly} that $\Delta^{\mathcal I}$ and $P^{\mathcal I}$ consist of $2\tau_0+1$ linearly independent polynomials. From Theorem~\ref{theo:linkombKE} we know that the linear combinations from \eqref{eq:linkombi} fulfill the key equation. Now, we show that \textit{any} pair of polynomials satisfying the key equation can be represented by these linear combinations.

Now, we rewrite the key equation \eqref{eq:keyequation} in order to prove that the linear combinations \eqref{eq:linkombi}, provide \textit{all} solutions of the key equation. $\Lambda(x)$ is a linearized polynomial with $q$-degree $\tau$ and hence has $\tau + 1$ unknown coefficients. Therefore, %If we carry out the symbolic multiplication of the key equation and order the result with the $q$-degree of $x$, we obtain:
\begin{equation*}
\begin{split}
\Omega(x)  = &\ \ \Lambda(S(x)) \mod x^{[d-1]}\\
= & \ \ x^{[0]}(\Lambda_0 S_0)\\
&+ x^{[1]}(\Lambda_0 S_1 + \Lambda_1 S_0^{[1]})\\
%&+ x^{[2]}(\Lambda_0 S_2 + \Lambda_1 S_1^{[1]} + \Lambda_2 S_0^{[2]})\\
&+ \dots\\
&+ x^{[\tau]}(\Lambda_0 S_{\tau} + \Lambda_1 S_{\tau-1}^{[1]}+ \dots +\Lambda_{\tau} S_0^{[\tau]})\\
&+ x^{[\tau+1]}(\Lambda_0 S_{\tau+1} + \Lambda_1 S_{\tau}^{[1]}+ \dots  +\Lambda_{\tau} S_1^{[\tau]})\\
&+ \dots\\
&+ x^{[d-2]}(\Lambda_0 S_{d-2} + \Lambda_1 S_{d-3}^{[1]}+ \dots  +\Lambda_{\tau} S_{d-\tau-2}^{[\tau]}).
\end{split}
\end{equation*}
%The modulo operation is included, since calculating modulo $x^{[d-1]}$ means cutting off after $q$-degree $d-2$. 
We claim in \eqref{eq:degprob} that $\deg_q \Omega(x)< \tau$ and hence the coefficients of $x^{[\tau]}, x^{[\tau+1]},\dots,x^{[d-2]}$ must be zero. Consequently, we have $d-2-\tau+1$ equations which fix some $\Lambda_i$. 

We have to check if these equations are linearly independent in order to know how many coefficients are fixed. If we rewrite this linear system of equations in matrix form, we obtain:
\begin{equation}\label{eq:systemS}
\mathbf S \cdot (\Lambda_{\tau}, \Lambda_{\tau-1}, \dots,\Lambda_{0})^T = \mathbf 0,
\end{equation}
where $\mathbf S$ is a $(d-\tau-1)\times(\tau+1)$ matrix:
\begin{small}
\begin{equation}\label{eq:matS}
\mathbf S = \left( \begin{array}{cccccc}
S_{0}^{[\tau]}&S_{1}^{[\tau-1]}&S_{2}^{[\tau-2]}&\dots&S_{\tau}\\
S_{1}^{[\tau]}&S_{2}^{[\tau-1]}&S_{3}^{[\tau-2]}&\dots&S_{\tau+1}\\
S_{2}^{[\tau]}&S_{3}^{[\tau-1]}&S_{4}^{[\tau-2]}&\dots&S_{\tau+2}\\
\vdots&\vdots&\vdots&\vdots&\vdots\\
%S_{d-\tau-3}^{[\tau]}&S_{d-\tau-2}^{[\tau-1]}&S_{d-\tau-1}^{[\tau-2]}&\dots&S_{d-3}\\
S_{d-\tau-2}^{[\tau]}&S_{d-\tau-1}^{[\tau-1]}&S_{d-\tau}^{[\tau-2]}&\dots&S_{d-2}\\[1ex]
\end{array}
\right).
\end{equation}
\end{small}
For this matrix, the following lemma holds. %, which has rank $\min\lbrace \wt(\mathbf e), d-1-\tau\rbrace$. If $\rank(\mathbf S) < d-1-\tau$ the solution is unique and this solution is given by Algorithm~\ref{algo:UpToHalfD}. 
% where $\mathbf S$ is the $(d-\tau-1)\times(\tau+1)$ matrix from Lemma~\ref{theo:rankS}, which has rank $\min\lbrace \wt(\mathbf e), d-1-\tau\rbrace$. If $\rank(\mathbf S) < d-1-\tau$ the solution is unique and this solution is given by Algorithm~\ref{algo:UpToHalfD}. 

\begin{Lemma}\label{theo:rankS}
Let $\mathbf S$ \eqref{eq:matS} be the matrix of syndrome elements satisfying the requirements of Problem~\ref{prob:dec}, 
%{\renewcommand{\arraystretch}{1.4}
% {\setlength\arraycolsep{0.23em}
% \begin{small}
% \begin{small}
% \begin{equation}
% \mathbf S = \left( \begin{array}{cccccc}
% S_{0}^{[\tau]}&S_{1}^{[\tau-1]}&S_{2}^{[\tau-2]}&\dots&S_{\tau}\\
% S_{1}^{[\tau]}&S_{2}^{[\tau-1]}&S_{3}^{[\tau-2]}&\dots&S_{\tau+1}\\
% S_{2}^{[\tau]}&S_{3}^{[\tau-1]}&S_{4}^{[\tau-2]}&\dots&S_{\tau+2}\\
% \vdots&\vdots&\vdots&\vdots&\vdots\\
% %S_{d-\tau-3}^{[\tau]}&S_{d-\tau-2}^{[\tau-1]}&S_{d-\tau-1}^{[\tau-2]}&\dots&S_{d-3}\\
% S_{d-\tau-2}^{[\tau]}&S_{d-\tau-1}^{[\tau-1]}&S_{d-\tau}^{[\tau-2]}&\dots&S_{d-2}\\[1ex]
% \end{array}
% \right),
% \end{equation}
% \end{small}
%}
then $\rank(\mathbf S)=\min\lbrace  d-1-\tau, t=\rank_q(\mathbf e)\rbrace$.

% If $\rank(\mathbf S) < d-1-\tau$ and $\tau >\left\lfloor \frac{d-1}{2}\right\rfloor$, then the error can be corrected by the Gabidulin Algorithm (Algorithm~\ref{algo:UpToHalfD}).
\end{Lemma}
\begin{IEEEproof}
We assume in Problem~\ref{prob:dec} that $t=\rank_q(\mathbf e) \leq \tau$. With \eqref{eq:decompe} and \eqref{eq:calcs}, we can rewrite the syndrome by (see \cite{Gabidulin_TheoryOfCodes_1985}):
\begin{equation}\label{eq:rewriteSynd}
\mathbf s = \mathbf E \cdot \mathbf Y \cdot \mathbf H^T=\mathbf E \cdot\mathbf X,
\end{equation}
where the elements of the $t \times (d-1)$-matrix $\mathbf X$ in row $i$ and column $j$ are: $\mathbf{X}_{i,j}=x_i^{[j-1]}$ for $i=1,\dots,t$ and $j=1,\dots,d-1$.
% \begin{equation}
% \mathbf X = \left( \begin{array}{cccc}
% x_1&x_1^{[1]}&\dots&x_1^{[d-2]}\\
% x_2&x_2^{[1]}&\dots&x_2^{[d-2]}\\
% \vdots&\vdots&\vdots&\vdots\\
% x_t&x_t^{[1]}&\dots&x_t^{[d-2]}\\
% \end{array}
% \right)
% \end{equation}
The $x_i$ are linearly independent over $\mathbb{F}_{q}$ as shown in \cite{Gabidulin_TheoryOfCodes_1985}. \eqref{eq:rewriteSynd} yields:
\begin{equation}
S_l^{[k]}=\sum \limits_{j=1}^{t}E_j^{[k]}x_j^{[k+l]},
\end{equation}
where $k$ is an arbitrary integer and $l=0,\dots,d-2$. We can then decompose $\mathbf S$ as follows:
{\begin{small}
\begin{multline}
\mathbf S = \mathbf{\hat X} \cdot \mathbf{\hat E} = 
\left( \begin{array}{ccccc}
x_1^{[\tau]}&x_2^{[\tau]}&\dots&x_{t}^{[\tau]}\\
x_1^{[\tau+1]}&x_2^{[\tau+1]}&\dots&x_{t}^{[\tau+1]}\\
\vdots&\vdots&\vdots&\vdots\\
x_1^{[d-2]}&x_2^{[d-2]}&\dots&x_{t}^{[d-2]}\\
\end{array}
\right)\\[1ex]
\cdot
\left(\begin{array}{ccccc}
E_1^{[\tau]}&E_1^{[\tau-1]}&E_1^{[\tau-2]}&\dots&E_1^{[0]}\\
E_2^{[\tau]}&E_2^{[\tau-1]}&E_2^{[\tau-2]}&\dots&E_2^{[0]}\\
\vdots&\vdots&\vdots&\vdots&\vdots\\
E_{t}^{[\tau]}&E_{t}^{[\tau-1]}&E_{t}^{[\tau-2]}&\dots&E_{t}^{[0]}\\ 
\end{array}
\right),
\end{multline}\end{small}}
where $x_1,\dots,x_t$ and $E_1,\dots,E_t$ are linearly independent over $\mathbb{F}_{q}$. Both matrices, $\mathbf{\hat X}$ and $\mathbf{\hat E}$, are linearized Vandermonde-like matrices. Such a matrix always has full rank \cite{Lidl-Niederreiter:FF1996}. Therefore, $\rank(\mathbf{\hat X})=\min\lbrace d-\tau-1,t\rbrace$ and $\rank(\mathbf{\hat E})=\min\lbrace t,\tau+1\rbrace=t$. Since the rows of $\mathbf{\hat E}$ are linearly independent,
\begin{equation}
\rank(\mathbf S) = \min\lbrace d-\tau-1,t=\rank_q(\mathbf e) \rbrace.
\end{equation}
% If $\rank(\mathbf S)<d-\tau-1$, then $\rank(\mathbf S)=\rank_q(\mathbf e) <d-\tau-1$. If $\tau >\left\lfloor \frac{d-1}{2}\right\rfloor$, this is equivalent to $\rank_q(\mathbf e) < \left\lfloor \frac{d-1}{2}\right\rfloor - \tau_0$. This error can always be corrected by Algorithm~\ref{algo:UpToHalfD}.
\end{IEEEproof}

% \begin{Lemma}
% Let $\mathbf{c}_1$ be the transmitted codeword and $\mathbf r= \mathbf{c}_1+\mathbf{e}_1$ is received. Let $\wt(\mathbf{e}_1)>\left\lfloor \frac{d-1}{2}\right\rfloor$ hold. Moreover, assume that a codeword $\mathbf{c}_2$ exists with $\mathbf r = \mathbf{c}_2+\mathbf{e}_2$, where $\wt(\mathbf{e}_2) \leq \left\lfloor \frac{d-1}{2}\right\rfloor$. The decomposition of $\mathbf S$ from Lemma~\ref{theo:rankS} works for both possible errors. 
% \end{Lemma}
% \begin{IEEEproof}
% We can decompose $\mathbf S$ in two ways:
% \begin{equation}
% \mathbf S = \mathbf{X}_1 \cdot \mathbf{E}_1 = \mathbf{X}_2 \cdot \mathbf{E}_2,
% \end{equation}
% as in Lemma~\ref{theo:rankS} with $t_1 = \wt(\mathbf{e}_1)$ and $t_2 = \wt(\mathbf{e}_2)$. We know that $\left\lfloor \frac{d-1}{2}\right\rfloor < t_1 \leq \tau$, hence:
% \begin{equation}
% \begin{split}
% \rank(\mathbf{X}_1)&=\min \lbrace t_1, d-\tau-1 \rbrace = d-\tau-1,\\
% \rank(\mathbf{E}_1)&=\min \lbrace t_1, \tau+1 \rbrace = t_1.
% \end{split}
% \end{equation}
% Thus, with this decomposition: $\rank(\mathbf S)=\min \lbrace t_1, d-\tau-1 \rbrace = d-\tau-1$. 
% 
% For the second decomposition we know that $t_2 \geq d- t_1$ as two codewords differ at least by $d$. As $t_1 \leq \tau$, we have: $t_2 \geq d-\tau > d- \tau -1$. Hence,
% \begin{equation}
% \begin{split}
% \rank(\mathbf{X}_2)&=\min \lbrace t_2, d-\tau-1 \rbrace = d-\tau-1,\\
% \rank(\mathbf{E}_2)&=\min \lbrace t_2, \tau+1 \rbrace = t_2.
% \end{split}
% \end{equation}
% Thus, also with this decomposition: $\rank(\mathbf S)=\min \lbrace t_2, d-\tau-1 \rbrace = d-\tau-1$.
% \end{IEEEproof}

\begin{Theorem}\label{theo:linkombi}
Let $\Delta^{\mathcal I}$ and $P^{\mathcal I}$ be the sets of $2 \tau_0 +1$ linearly independent polynomials calculated with Algorithm~\ref{algo:BeyondHalfD}. 
%Let $\tau = \left\lfloor \frac{d-1}{2}\right\rfloor + \tau_0$, $\tau_0 \geq 0$ and $\tau < d-1$ hold.
Let $\left\lfloor \frac{d-1}{2}\right\rfloor< \tau < d-1$ and $\rank_q(\mathbf e) \leq \tau$ hold.

Any pair of polynomials $\lbrace \Lambda(x), \Omega(x) \rbrace$ satisfying the key equation \eqref{eq:keyequation} with the degree constraints \eqref{eq:degprob}, can be calculated by the linear combinations of the polynomials from $\Delta^{\mathcal I}, P^{\mathcal I}$ given in \eqref{eq:linkombi}. That means, the sets $\Delta^{\mathcal I}, P^{\mathcal I}$ are a basis for all solutions of the key equation \eqref{eq:keyequation}.
\end{Theorem}
\begin{IEEEproof}
In the following, we show that only $2\tau_0 +1$ coefficients of $\Lambda(x)$ can be chosen arbitrarily, if \eqref{eq:keyequation} has to be fulfilled. Hence, the set of $(q^m)^{2\tau_0+1}$ different linear combinations from \eqref{eq:linkombi} constitutes the set of all possible $\Lambda(x)$.% satisfying the key equation.

% We can rewrite \eqref{eq:keyeqtheo} by:
% \begin{equation}
% \Omega(x) = \Lambda(S(x)) \mod x^{[d-1]},
% \end{equation}
% $\Lambda(x)$ is a linearized polynomial with $q$-degree $\tau$ and hence has $\tau + 1$ unknown coefficients. If we carry out the symbolic multiplication of the key equation and order the result with the $q$-degree of $x$, we obtain:
% \begin{equation*}
% \begin{split}
% \Omega(x)  = &\ \ \Lambda(S(x)) \mod x^{[d-1]}\\
% = & \ \ x^{[0]}(\Lambda_0 S_0)\\
% &+ x^{[1]}(\Lambda_0 S_1 + \Lambda_1 S_0^{[1]})\\
% %&+ x^{[2]}(\Lambda_0 S_2 + \Lambda_1 S_1^{[1]} + \Lambda_2 S_0^{[2]})\\
% &+ \dots\\
% &+ x^{[\tau]}(\Lambda_0 S_{\tau} + \Lambda_1 S_{\tau-1}^{[1]}+ \dots +\Lambda_{\tau} S_0^{[\tau]})\\
% &+ x^{[\tau+1]}(\Lambda_0 S_{\tau+1} + \Lambda_1 S_{\tau}^{[1]}+ \dots  +\Lambda_{\tau} S_1^{[\tau]})\\
% &+ \dots\\
% &+ x^{[d-2]}(\Lambda_0 S_{d-2} + \Lambda_1 S_{d-3}^{[1]}+ \dots  +\Lambda_{\tau} S_{d-\tau-2}^{[\tau]}).
% \end{split}
% \end{equation*}
% The modulo operation is included, as calculating modulo $x^{[d-1]}$ means cutting off after $q$-degree $d-2$. In \eqref{eq:degconstrke} we claim that $\Omega(x)$ has $\deg_q < \tau$ and hence the coefficients of $x^{[\tau]}, x^{[\tau+1]},\dots,x^{[d-2]}$ have to be zero. Consequently, we have $d-2-\tau+1$ equations which fix some $\Lambda_i$. 

With \eqref{eq:systemS}, some coefficients of $\Lambda(x)$ are fixed. Lemma~\ref{theo:rankS} shows that $\rank(\mathbf S) =\min\lbrace d-1-\tau, t=\rank_q(\mathbf e)\rbrace$. 

If $\rank(\mathbf S)<d-\tau-1$, then $\rank(\mathbf S)=\rank_q(\mathbf e) <d-\tau-1$. We assume in the theorem $\tau >\left\lfloor \frac{d-1}{2}\right\rfloor$, hence this is equivalent to $\rank_q(\mathbf e) < \left\lfloor \frac{d-1}{2}\right\rfloor$ and this error can always be corrected by Algorithm~\ref{algo:UpToHalfD}. %This case can be recognized during the SEEA with Lemma~\ref{theo:ranksmall} and this error can always be corrected by Algorithm~\ref{algo:UpToHalfD}. %In practice, we break Algorithm~\ref{algo:BeyondHalfD} if Lemma~\ref{theo:ranksmall} is fulfilled. This is equivalent to using Algorithm~\ref{algo:UpToHalfD}.
Otherwise, $\rank(\mathbf S)=d-1-\tau$ and that means that $d-1-\tau$ coefficients of $\Lambda(x)$ are fixed. Since the number of coefficients of $\Lambda(x)$ is $\tau+1$, the number of free coefficients is:
\begin{equation}
\tau+1-(d-1-\tau)=(2\tau-(d-1))+1 = 2\tau_0+1.
\end{equation}
Hence, there are only $(q^m)^{2\tau_0+1}$ possible $\Lambda(x)$. They can be calculated by the $(q^m)^{2\tau_0+1}$ linear combinations from \eqref{eq:linkombi} as the linear combinations satisfy the key equation and fulfill the degree constraints. Equivalently, all $\Omega(x)$ can be calculated by the linear combinations from \eqref{eq:linkombi}.
\end{IEEEproof}

% \begin{Lemma}\label{theo:ranksmall}
% $\rank_q(\mathbf e) < \left\lfloor \frac{d-1}{2}\right\rfloor - \tau_0$, if and only if there exists a pair $\lbrace R_i(x),U_i(x) \rbrace$ calculated by Algorithm~\ref{algo:BeyondHalfD} with $\deg_q R_i(x) < \deg_q U_i(x) < \left\lfloor \frac{d-1}{2}\right\rfloor - \tau_0$.
% \end{Lemma}
% % \begin{IEEEproof}
% % \end{IEEEproof}
% This proof was skipped due to place restrictions.

The proof of Theorem~\ref{theo:linkombi} is one of the main results of this paper as it can be done in a similar way for RS codes. This proof is more descriptively than the proof given in \cite{IRS_Bossert} for RS codes. %and no special properties of the polynomials obtained with the (S)EEA as in \cite{IRS_Bossert} have to be used.

\section{Conclusion}\label{sec:concl}
% generalization of \cite{IRS_Bossert}, if $m=1$ then $\alpha^{[i]}=\alpha$ which corresponds to the linear case (RS codes)
In this paper, we have presented an efficient algorithm that provides a basis for all solutions of the key equation for decoding of $\mathcal{G}$-codes up to a certain radius $\tau$. This algorithm requires $\mathcal O(\tau^2)$ operations in $\mathbb{F}_{q^m}$ and can be applied for decoding of $\mathcal{G}$-codes beyond half the minimum rank distance. Our algorithm is based on a symbolic equivalent of the EA and is a generalization of the BBA. 
% \section*{Acknowledgment}
% This work was supported by the German Research Council (DFG) under Grant No. Bo867/21-1.

%\IEEEtriggeratref{0}
% \bibliographystyle{IEEEtran}
% \bibliography{antoniawachter}

\end{document}